\documentclass{style/nseJournal}

\begin{document}

\title{Hybrid simulations of FRC merging and compression} %title of paper

% Use the \addAuthor macro to add authors in the order they should appear. The second argument corresponds to
% the affiliation declared below.
% The corresponding author should be wrapped in \correspondingAuthor
\addAuthor{\correspondingAuthor{E. V. Belova}}{a}
% The corresponding author's email can be specified using \correspondingEmail
\correspondingEmail{ebelova@pppl.gov}
\addAuthor{S. E. Clark}{b}
\addAuthor{R. Milroy}{b}
\addAuthor{G. Votroubek}{b}
\addAuthor{A. Pancotti}{b}
\addAuthor{D. Kirtley}{b}

% Affiliations can be added in the order they should appear. For breaks in addresses, use either \\ or \tabularnewline
\addAffiliation{a}{Princeton Plasma Physics Laboratory, Theory Department\\ P.O. Box 451, MS29, Princeton, New Jersey 08543}
\addAffiliation{b}{Helion Energy Inc., 1415 75th St SW, Everett, WA 98203}

% Add keywords to appear in Abstract in the order they should appear
\addKeyword{FRC}
\addKeyword{Plasma physics}
\addKeyword{Reconnection}
\addKeyword{Plasma compression}

\titlePage

% ===========DEFINITIONS
\def\alt{\mathrel{\rlap{\lower3.5pt\hbox{$\mathchar"218$}}\raise 2pt
                                           \hbox{$\mathchar"13C$}}}
\def\agt{\mathrel{\rlap{\lower3.5pt\hbox{$\mathchar"218$}}\raise 2pt
                                           \hbox{$\mathchar"13E$}}}
\def\oci{\hbox{$\omega_{ci}$}}
\def\ooci{\hbox{$\omega/\omega_{ci}$}}
\def\dbp{\hbox{$\delta B_\|$}}
\def\dbperp{\hbox{$\delta B_\perp$}}
\def\dbb{\hbox{$\delta {\bf B}$}}
\def\Alfven{\hbox{Alfv\'en }}
\def\En{\hbox{$\mathcal{E}$}}
%%%%%%%%% END DEFINITIONS

\begin{abstract}
An improved understanding of Field Reversed Configuration (FRC) merging and stability in high acceleration and compression magnetic fields is needed to speed up the development of the pulsed fusion concept developed at Helion Energy. All previous theoretical and simulation work on FRC merging and compression was performed using 2D MHD models. The results of novel 2D hybrid simulations (fluid electrons and full-orbit kinetic ions) of FRC  merging and compression are presented. Results of kinetic 
 and MHD simulations, computed using the HYM code, are compared and analyzed.  In cases without axial magnetic compression, both the MHD and hybrid simulations show a high sensitivity to the initial parameters (i.e. FRC separation, velocity, normalized separatrix radius, and plasma viscosity), showing that FRCs with large elongation and separatrix radius either do not merge or merge partially, forming a doublet FRC. Application of a mirror coil field at the FRC ends with increasing strength is shown to lead to fast and complete merging of the FRCs in MHD and kinetic simulations.
\end{abstract}

\section{Introduction}
\label{sec:intro}

Cylindrical, high-field, pulsed solenoidal coils form the basis of the fusion platform being developed at Helion Energy. The geometrically simple, compact, and high beta ($\beta \sim 1$) Field Reversed Configuration (FRC) is an excellent target for compression with such coils. With external fields that can potentially be pulsed up to ~ 20 T, a compressed 10 keV FRC would have a peak density of $10^{23}~m^{-3}$. Extrapolating from previously derived empirical FRC confinement scaling relationships, the energy gain for this FRC with 20 mWb of internal, poloidal flux, would be $Q \sim 2$. Helion’s sixth generation prototype, "Trenta", was a double ended machine, where each end formed an FRC plasmoid and accelerated it to high velocity ($\mathcal{M_A} \sim 1$), while simultaneously compressing it. The two plasmoids collided and merged in a compression chamber in the center of the machine, where the kinetic energy of the FRCs was rapidly converted to ion thermal energy. The magnetic field in the compression section then ramps up to confine and adiabatically heat the newly merged FRC. In Trenta, FRCs were formed and compressed to above 9 keV plasma temperatures at peak magnetic fields of 8 T, inferred from magnetic flux diagnostics and interferometry. The experimental results from Trenta set new records for FRC temperature and density. The fusion performance demonstrated in Trenta still needs improvement for Helion’s fusion technology to be a viable source of clean energy for the grid. Helion is building an increased scale fusion prototype with the goal of demonstrating electricity from fusion. An improved understanding of FRC merging and stability in high acceleration and compression fields will speed up the development of this technology.

A primary challenge in confining and heating FRC plasmas is managing low toroidal mode number global instabilities. Specifically, the $n=1$ tilt mode and the $n=2$ rotational mode are the most challenging unstable modes in the conventional quasi-steady state FRCs \cite{T88}. These instabilities are either frequently observed to grow and terminate the FRCs (the $n=2$ rotational mode) or thought to be responsible for difficulties of forming an FRC with large $S^*/E$ ratio (the $n=1$ tilt mode) \cite{T88a, B03b}. Here $S^*=R_s/d_i$ is the FRC kinetic parameter defined as ratio of the separatrix radius to the ion collisionless skin depth, and $E=Z_s/R_s$ is the separatrix elongation defined as ratio of the separatrix half-length to its radius.

Tilt mode in FRC plasmas is strongly unstable in the MHD model with $\gamma \sim v_A/Z_s$, but its growth rate is reduced by the kinetic effects due to large Larmor radius of the thermal ions \cite{T88,S11}. The kinetic stabilization of the tilt mode can be described by an empirical scaling law which provides the growth rate as a function of the S*/E parameter \cite{B03b}. Since the MHD predictions are in contrast with experimental results, significant efforts have been deployed to include kinetic effects in theoretical models and in numerical simulations \cite{B00, B04}. 

FRC spin up due to particle losses with preferential azimuthal angular momentum \cite{E71, B06a} and “end shorting” of open magnetic field lines on conducting structures \cite{T88, M07} may trigger the $n=2$ instability when the FRC rotation velocity exceeds the ion diamagnetic velocity \cite{S79a, H83, B14b}. Experimentally the $n=2$ rotational mode can be stabilized by end biasing \cite{T12} or application of a quadrupole field \cite{T88}. Numerical simulation of the FRC spin-up and growth of the $n=2$ rotational instability requires kinetic description of the thermal ions \cite{B04, B06a}.

Merging of FRCs is accompanied by a conversion of plasma kinetic energy and the magnetic energy into heat, resulting in the high-beta configurations with finite flows. The highly dynamical nature of merging, which involves reconnection of magnetic field lines, allows the plasma rapid access to a particular minimum-energy state. Experimental FRC parameters, such as values of $x_s$ ($x_s=R_s/R_c$ – ratio of the separatrix radius to the flux conserver radius), the line-tying and finite plasma pressure at the separatrix can have an effect on the merging and relaxation process. In some cases, merging can be incomplete, resulting in doublet type of FRC with two null points or the FRCs bouncing. Experimental diagnostics for very hot, merging, and dynamical plasmas can be very challenging. On the other hand, understanding the physics of the FRC merging, its dependence on the plasma parameters, and scaling with the device size is important for achieving the goals of the Helion experimental program and planning of the next-step fusion prototype.

In this paper initial results of FRC merging and compression simulations are presented.
The 2D hybrid simulations (full orbit kinetic ions and fluid electrons) using HYM code \cite{B00} have been performed and compared with the MHD simulations for the same plasma and external field parameters.
The paper is organized as follows: Section~\ref{sec:simulation} describes the numerical model used in this study. 
Section~\ref{sec:results} presents the simulation results:
Subsection~\ref{subsec:A} describes FRC merging in a fixed external field for FRCs with different parameters. 
Subsection~\ref{subsec:B} describes FRC merging under compression.
Discussion and conclusions are given in Sec.~\ref{sec:conclusion}.

\section{Simulation Approach}
\label{sec:simulation}
In the MHD model, the FRCs are known to be unstable to a set of different global modes including the kink-type modes with either axial or radial polarization (in terms of dominating velocity perturbation) and interchange-type modes \cite{B00} with MHD growth rates increasing with the toroidal mode number, $n$. However, the experiments tend to operate in the kinetic regimes, which leads to a strong ion FLR stabilization of high-$n$ modes. As a result, advanced numerical models, including thermal plasma large-Larmor-radius effects and kinetic effects are needed for realistic modeling of FRCs. Comparison of the results of 2D hybrid and MHD simulations of counter-helicity spheromak merging \cite{K11,S15,B16}, for example, shows that even in the MHD-like regime, there are significant differences between the kinetic and MHD models, and demonstrate the need for a kinetic description of the merging plasmas. 

In this paper we use the 3D nonlinear hybrid code (HYM) to study 2D FRC merging with and without axial compression\cite{B00}. 
The hybrid version of HYM, which uses a full-orbit kinetic description for thermal ions and a fluid description of the electrons, has been used in this study.
Additionally, a single-fluid, resistive MHD version of the HYM code is used to model FRC merging for the same parameters and initial conditions.
Comparisons between the MHD results and kinetic physics models is used to evaluate the importance of two-fluid and ion kinetic effects in FRC merging. 
HYM was initially developed for FRC studies and has been extensively benchmarked and validated for FRC modeling applications \cite{B04,G06,M11} as well as for NSTX(-U) experiments \cite{B19, F17}. HYM has also been used to model 3D spheromak merging using both the MHD and kinetic version of the code \cite{M11, B16}.

Initial conditions for the FRC merging simulations have been generated by solving the Grad-Shafranov equation for a single FRC assuming the scalar plasma pressure and zero initial ion rotation.
This equilibrium configuration was then mirror flipped in the axial direction to generate an axially symmetric configuration with two FRCs.
In order to avoid the possibility of current density singularities at the midplane, the poloidal flux was then re-calculated by solving the Grad-Shafranov equation with the right-hand-side replaced with $RJ_\phi$, where the toroidal current $J_\phi$ is the current density of the two FRCs. 
Perfectly conducting boundary conditions have been applied in simulations without axial compression, and time dependent boundary conditions have been used in the compression simulations by specifying a time-dependent and nonuniform profile of the toroidal component of vector potential $A_\phi$ at the simulation boundary $R=R_c$. 

In this paper the following normalization is used for all plasma parameters: pressure is normalized to $B_0^2/4\pi$, where $B_0$ is the initial value of the external magnetic
field at the FRC separatrix, density is normalized by the O-point density $n_0$ - initial peak value at the magnetic null, velocity is normalized to the \Alfven velocity defined as $v_A = B_0/\sqrt{4\pi n_0 m_i}$, length is normalized by $d_i$, the ion colissionless skin depth calculated for density value $n_0$, and time is normalized by $1/\oci$. 
The \Alfven time is defined as $t_A=R_c/v_A$, where $R_c$ is the flux conserver radius.
In the following simulations, FRC parameters ($S^*$, $E$, $x_s$ and separatrix $\beta$) and profiles has been varied to study their effects on the FRC merging.

\section{Simulation Results}
\label{sec:results}

\subsection{Merging without compression}
\label{subsec:A}
Initial 2D merging simulations have been performed for the following set of FRC parameters:
kinetic parameter $S^*=R_s/d_i=25.6$, elongation $E=Z_s/R_s=2.9$, $x_s=R_s/R_c=0.69$, plasma resistivity and viscosity corresponding to Lundquist number $S=R_cv_A/\eta=4600$ and Reynolds number $Re=R_cv_A/\nu=1900$  respectively.
The two FRCs were placed at a distance of $\Delta Z=180$ between their respective magnetic nulls at t=0.
In order to induce the FRCs to merge, the simulations either initialized the FRCs with an initial axial velocity towards the midplane, or applied a fixed magnitude mirror magnetic field at the flux conserver ends at t=0.

\begin{figure}[t]%[!htbp]
  \centering
  \includegraphics[width=3.2in]{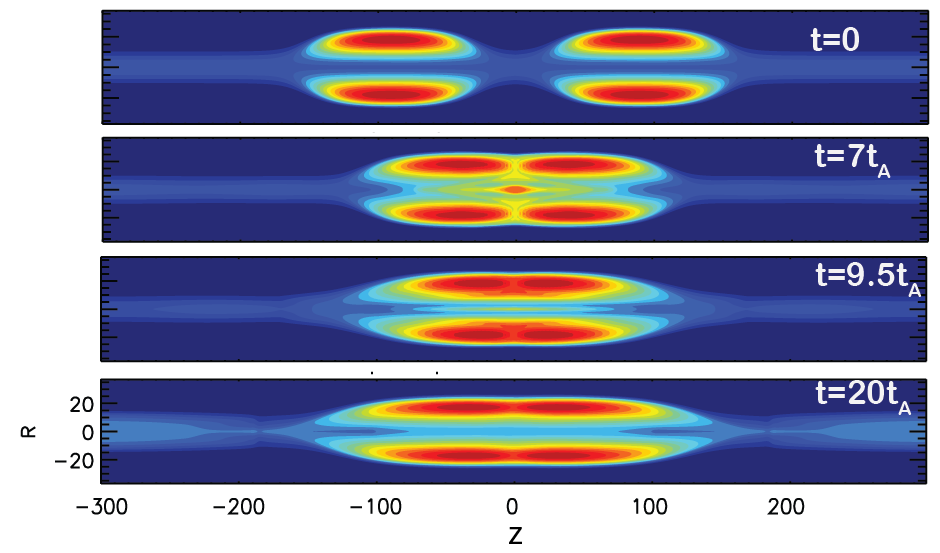}
  \caption{Contour plots of plasma pressure from 2D MHD simulations of FRC merging shown at different times $t=0 - 20t_A$. The peak values of pressure are $p_0=0.54$ ($t=0$) and 0.47 ($t=20t_A$), and the contours are equally spaced.}
  \label{fig:mhd1}
\end{figure}

Figure~\ref{fig:mhd1} shows contour plots of plasma pressure from MHD simulation of FRC merging at different times $t=0-20t_A$.
The initial velocity of $V_z=\pm 0.2v_A$ was added to the plasma inside the separatrix region at $t=0$.
As Fig.~\ref{fig:mhd1} shows, the FRCs merged partially, forming a doublet configuration with a final magnetic nulls separation $\Delta Z \approx 40$ and elongation increased to $E \sim 7$.
The peak reconnection happens at $t \sim 5t_A$.
Additional simulations have shown that the similar results (i.e. partial merging and the same final value of $\Delta Z$) have been obtained for a range of initial FRC velocities $|V_z|= 0.05- 0.25v_A$ or when a fixed mirror-field at the ends of the simulation region was added with mirror ratio 1.5.
%fig1
%

\begin{figure}
\centering
  \includegraphics[width=3.2in]{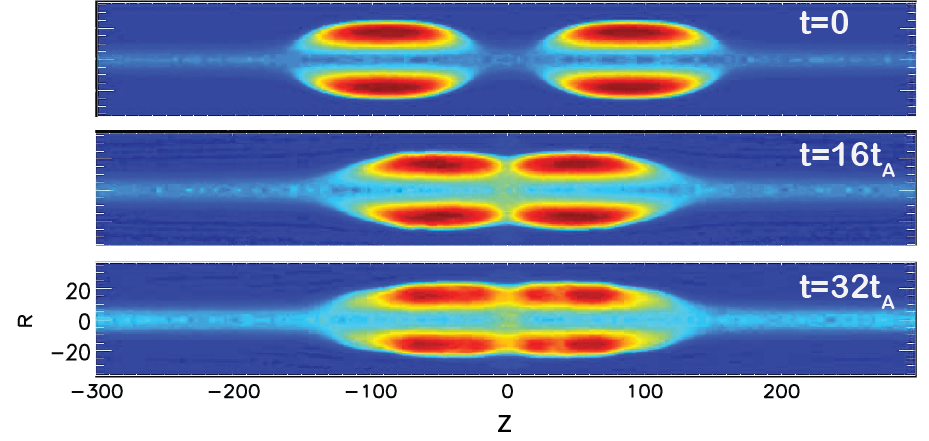}
  \caption{\label{fig:hybrid1} Contour plots of plasma density from 2D hybrid simulations of FRC merging shown at different times $t=0 - 32t_A$. }
\end{figure}

Results of hybrid simulations for the same initial FRC parameters and separation are shown in Fig.~\ref{fig:hybrid1}, where the contour plots of ion density are shown at $t=0-32t_A$.
In this case, no initial velocity was added, instead a mirror-field at the ends of the simulation region was turned on at $t=0$ with the mirror ratio 1.5.
Hybrid simulations show that for a relatively large value of $S^*$, the kinetic FRC merging is similar to that in the MHD model, and the FRCs do not merge completely.
The final configuration in hybrid simulations is a doublet FRC with larger value of final magnetic nulls separation $\Delta Z \approx 90$, but the total elongation comparable to that from the MHD simulations $E \sim 6.5$.
The separatrix radius after the merging ceased was comparable to the initial separatrix radius both in the hybrid and the MHD simulations.
These kinetic simulations do not show significant additional toroidal ion spin-up other than the spin-up to $v_\phi \sim 0.1-0.2v_A$ related to resistive flux decay and the associated particle loss from the closed field line region \cite{B06a}.
The resistive decay and spin-up were relatively small due to large value of Lundquist number and a short merging time.

%fig.2
%

Additional simulations have been carried out in order to study the dependence of the FRC merging on the configuration parameters.
In the simulations shown in Fig.\ref{fig:mhd1} and~\ref{fig:hybrid1} the FRCs have a relatively small initial separatrix beta $\beta_s=0.2$.
Simulations have also been performed for the same FRC elongation $E \sim 3$ and keeping the same value of $x_s=0.69$, but increasing the separatrix beta to $\beta_s=0.3$ (that required the change in the pressure profile to a more peaked one to satisfy the so-called average beta condition $\langle \beta \rangle=1-x_s^2/2$ \cite{B79}).
It was found that for larger values of $\beta_s$ the FRC do not merge even partially, but bounce off the midplane and move apart.
That behavior can be explained by a larger plasma density on the open field lines between the FRCs, since for a uniform initial plasma temperature, the separatrix density is proportional to separatrix beta $n_s/n_0=\beta_s$.

Simulations demonstrate even stronger dependence of the FRC merging on other initial parameters.
In particular, the merging is found to be very sensitive to values of the relative separatrix radius, $x_s$, and the initial separation of the FRCs.
In case of the MHD simulations, the results are also sensitive to the value of
plasma viscosity.

\begin{figure}
\centering
  \includegraphics[width=3.in]{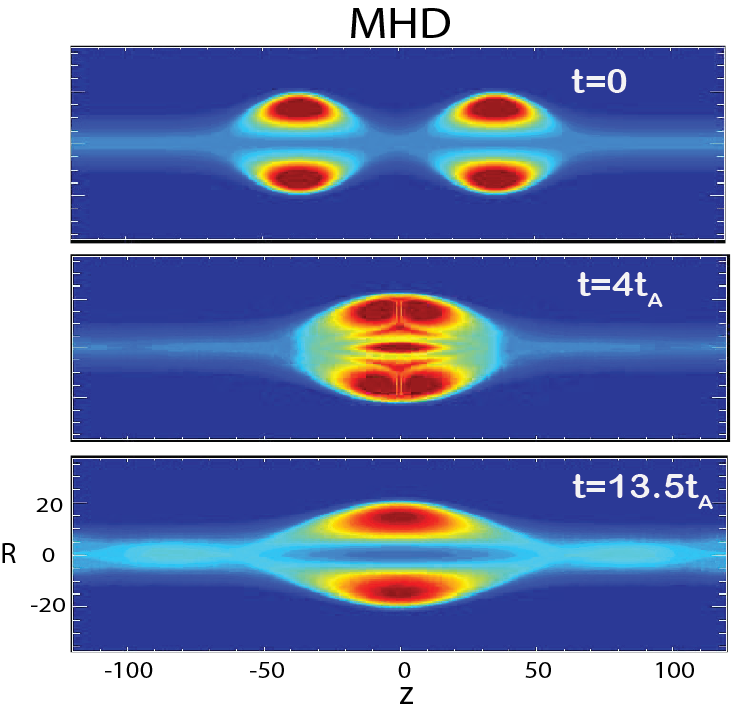}
  \caption{\label{fig:mhd2} Contour plots of plasma pressure from 2D MHD simulations of FRC merging for $x_s=0.53$, $E=1.5$ and $\beta_s=0.2$. The peak values of pressure are $p_0=0.89$ ($t=0$) and 0.64 ($t=13.5t_A$), the contours are equally spaced. }
\end{figure}

\begin{figure}
\centering
  \includegraphics[width=3.2in]{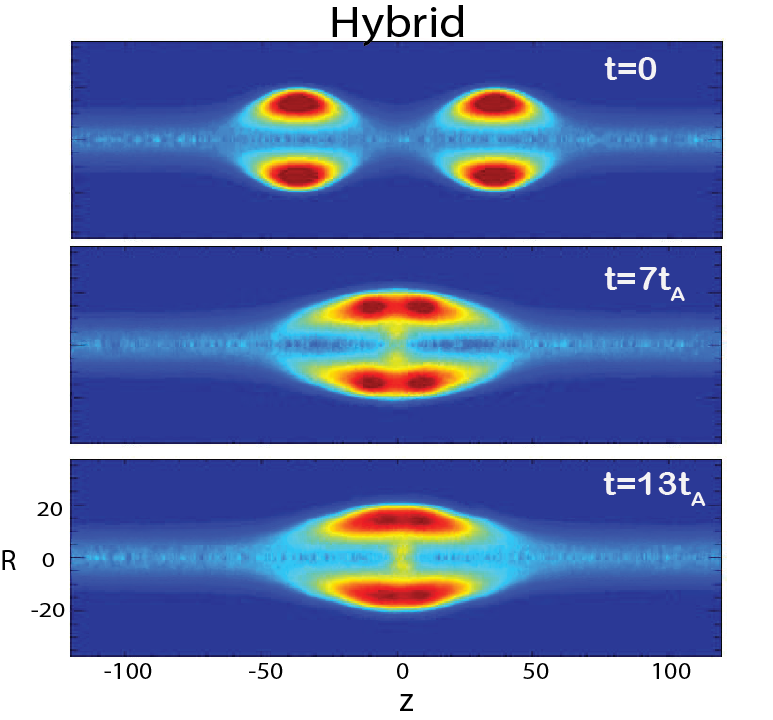}
  \caption{\label{fig:hybrid2} Contour plots of plasma density from 2D hybrid simulations of FRC merging for $x_s=0.53$, $E=1.5$ and $\beta_s=0.2$ shown at different times $t=0, 7, 13t_A$. }
\end{figure}

Figure~\ref{fig:mhd2} and Fig.~\ref{fig:hybrid2} show results of MHD and hybrid simulations of FRC merging for smaller initial values of the separatrix radius and elongation: $x_s=0.53$ and $E=1.5$ for $S^*=20$. 
%fig.3
%
%
%fig.4
%
These simulations were performed for the initial FRC separation $\Delta Z \approx 75$ and $V_z=\pm 0.1v_A$ at $t=0$.
In the MHD case, the FRCs merge completely by $t\sim 5t_A$, whereas in hybrid simulations, the merging is nearly complete on a somewhat longer time scale $t\sim 6-7t_A$.
The elongation increased in both cases to $E\sim 2.5$.
It is interesting to note that the final separatrix radius does not differ significantly from its initial value.
Since the change in the peak density was relatively small (within 20\%), the FRC kinetic parameter $S^*$ remains approximately the same as for the initial FRCs.
On the other hand, the separatrix beta increased from $\beta_s=0.2$ ($t=0$) to $\beta_s\approx 0.32$ ($t=13t_A$) and the pressure profile became less peaked.

\begin{figure}
\centering
  \includegraphics[width=3.in]{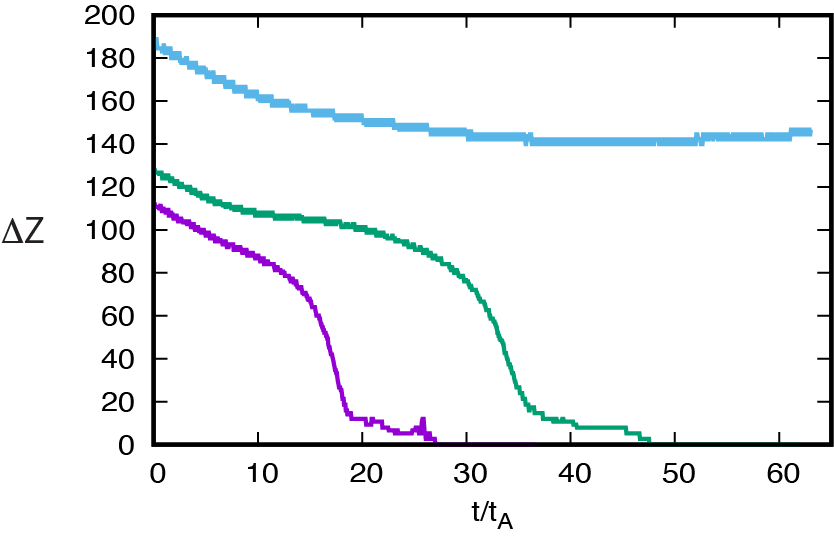}
  \caption{\label{fig:mhd_sep} Time evolution of FRCs separation from three simulations of FRC merging starting with different initial separation. }
\end{figure}

Additional simulations have been performed for the same FRC parameters as in Figs.~\ref{fig:mhd2} and~\ref{fig:hybrid2} but with larger initial separations (Fig.~\ref{fig:mhd_sep}).
Two simulations with $\Delta Z =110$ and $\Delta Z=125$ ($V_z=0.05v_A$ at t=0) 
still show a complete FRC merging, but on a much longer time scales: $t\sim 18t_A$ and $t\sim 36t_A$ respectively  as shown in Fig.~\ref{fig:mhd_sep} (purple and green lines).
%fig.5
%
Therefore a 14\% increase in the initial separation from $\Delta Z =110$ to $\Delta Z=125$ nearly doubled the merging time for the identical values of all other initial parameters.
Further increase in the separation to $\Delta Z \approx 185$  resulted in no merging.
In this case, the FRCs initially moved to the midplane to $\Delta Z \approx 140$ by $t\sim 50t_A$, and then started to slowly drift apart (shown in Fig.~\ref{fig:mhd_sep}, blue line).

The observed sensitivity of the merging results to initial parameters, in particular $\beta_s$, $x_s$ and the initial separation can be explained by the finite plasma density on the open field lines.
For two FRCs to meet at the midplane, the plasma present between the FRCs initially has to be squeezed past the FRCs toward the ends of the simulation region.
If the energy required to move this plasma out the the way is larger than the energy that could be released by the FRCs merging (plus the FRC initial kinetic energy), the merging becomes energetically prohibitive.
Both larger initial separation and larger values of $\beta_s$ increase the net mass of plasma to be displaced.
On the other hand, the large values of the separatrix radius, $x_s$, correspond to a reduction of the gap between the FRC and the cylindrical boundary $\sim R_c-R_s$, therefore increasing the effective viscosity and impeding the FRCs motion through the open-field-line plasma.

\subsection{Compressional merging}
\label{subsec:B}

Mirror field coils with gradually increasing current can be used to push, merge and compress the FRCs.
Time dependent boundary conditions have been implemented in the HYM code to model the FRC merging and compression.
The compression profile was used as $\delta A_\phi(z,R=R_c) \sim 0.5(1-\cos(\pi z/Z_c)) f(t)$, corresponding to increase of poloidal flux at the ends of the simulation region.
The time evolution was taken to be $f(t) \sim (1-\cos(\pi t/T))$, where $T=19t_A$ is the compression time.
By the end of compression the magnetic field magnitude at the $z=\pm Z_c$ was increased to $1.5B_0$, where $B_0$ is the initial magnitude of the external field at the FRC separatrix.

\begin{figure}
\centering
  \includegraphics[width=3.2in]{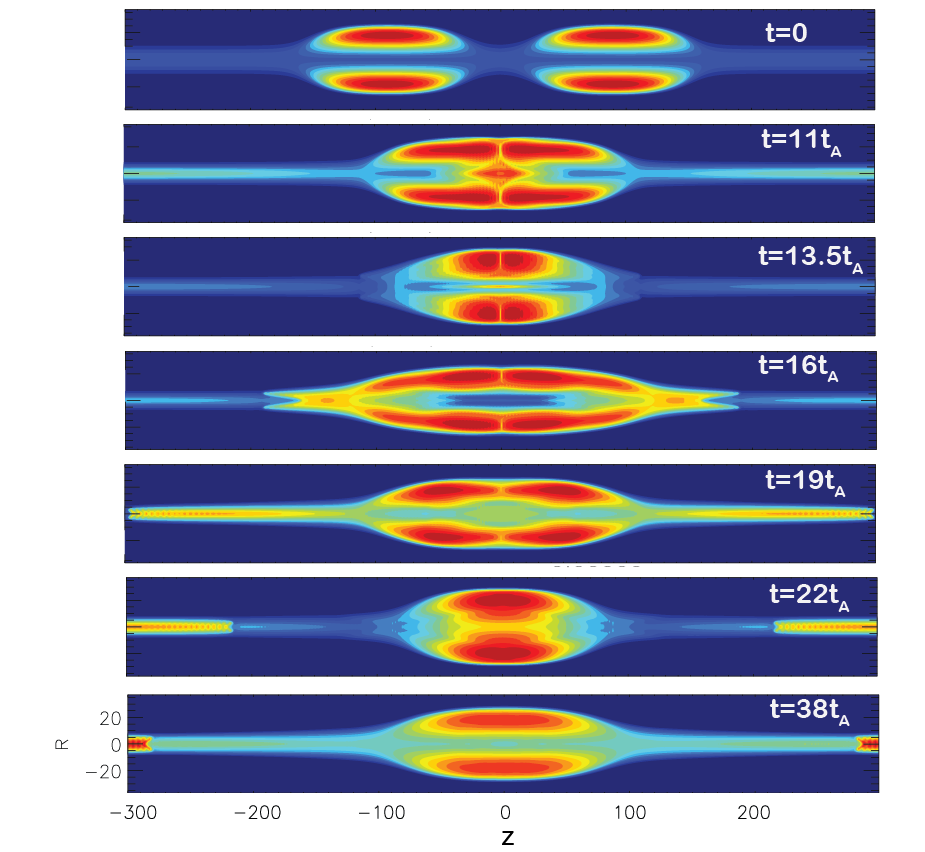}
  \caption{\label{fig:mhd3} Contour plots of plasma pressure from 2D MHD simulations of FRC merging and compression. Initial FRC parameters are the same as in Fig.~\ref{fig:mhd1}. The peak values of pressure are $p_0=0.54$ ($t=0$) and 0.71 ($t=38t_A$), the contours are equally spaced. }
\end{figure}

Figure~\ref{fig:mhd3} displays the results of MHD simulations of the FRC merging and compression.
%fig.6
%
Contour plots of plasma pressure are shown, where the peak values of pressure are $p_0=0.54$ ($t=0$) and 0.71 ($t=38t_A$). 
The initial FRC parameters are the same as in Figs.~\ref{fig:mhd1} and~\ref{fig:hybrid1}. 
The FRCs are accelerated by an increasing magnetic mirror field to axial velocity at the magnetic null $V_z \sim 0.4v_A$ at $t\approx 10t_A$, at which point the FRCs crash into each other with the separation, $\Delta Z$, deceasing sharply at $t\approx 11t_A$.
As the FRCs merge partially, the configuration becomes over-compressed axially, its separatrix radius increasing to the peak value $R_s \approx 29.4$.
Then the FRC began to expand axially, its separatrix radius drops to $R_s \approx 21$ and the elongation increases to $E\agt 7$.
During the ensuing axial oscillations, some of the plasma gets ejected from the closed field line region near the FRC x-points, as can be seen in Fig.~\ref{fig:mhd3} at $t=22t_A$.
After a couple of oscillations, the merging is complete, forming a single null FRC.
The final configuration has the separatrix radius comparable to that of the initial FRCs, larger elongation $E\sim 5$ and larger peak plasma pressure.

\begin{figure}
\centering
  \includegraphics[width=3.2in]{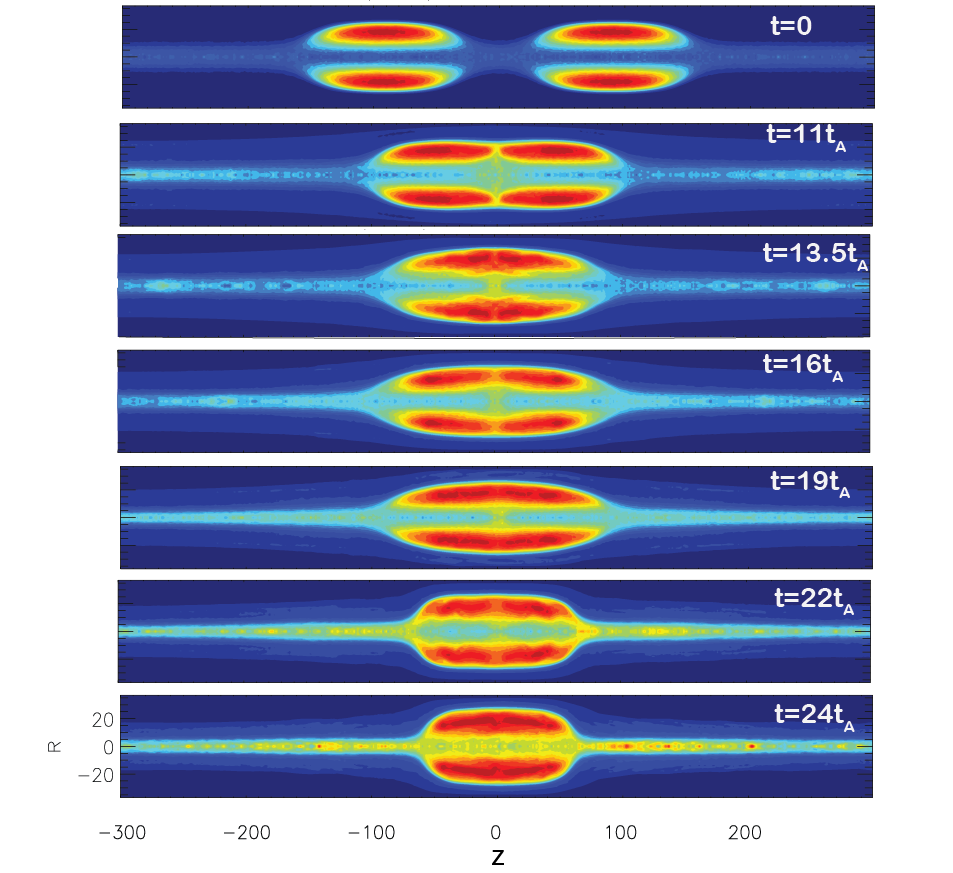}
  \caption{\label{fig:hybrid3} Contour plots of ion density from 2D hyrid simulations of FRC merging and compression. Initial FRC parameters are the same as in Figs.~\ref{fig:mhd1} and~\ref{fig:hybrid1}. Compressing field profile is the same as in Fig.\ref{fig:mhd3}. }
\end{figure}

\begin{figure}
\centering
  \includegraphics[width=3.2in]{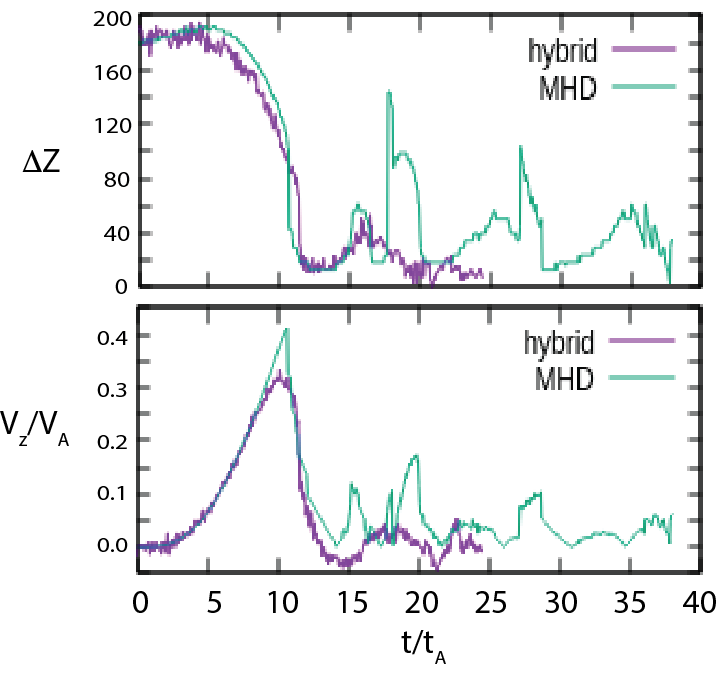}
  \caption{\label{fig:mhd_hybrid} Time evolution plots of FRCs separation and the magnitude of the axial velocity from hybrid (purple lines) and MHD  (blue-green lines) simulations. }
\end{figure}

For comparison, numerical results from the hybrid simulations of FRCs merging under compression are presented in Fig.~\ref{fig:hybrid3}, where the contour plots of ion density are shown at $t=0-24t_A$.
%fig.7
%
The global dynamics of the kinetic merging is very similar to that from the MHD simulations up to $t\sim 15t_A$.
This can be seen from the time evolution plots of FRCs separation and the magnitude of the FRC axial velocity shown in Fig.~\ref{fig:mhd_hybrid}.
%fig.8
%
Here the hybrid simulation results are shown by purple lines, and MHD results are shown by the blue-green lines.
The evolution of the FRC separation and axial velocity closely tracks those from the MHD simulations till $t\sim 15t_A$.
After the initial approach, the FRCs merge partially at $t\sim 10-16t_A$.
However, the amplitude and frequency of the subsequent axial oscillations are reduced compared to the MHD case.
Analysis of the local magnetic field structure near the midplane (not shown) also shows the differences between the kinetic and the MHD merging.
Thus, the magnetic field reconnection in the kinetic case is accompanied by the generation of the transient antisymmetric toroidal field (quadrupole magnetic field typically associated with the Hall reconnection).
In addition, the comparison of results from 2D hybrid simulations and
MHD simulations shows thicker and shorter current layer in hybrid simulations.  Kinetic simulations also show almost an order-of-magnitude lower radial outflow velocities and much wider ion velocity profiles near the reconnection region.
The somewhat reduced axial oscillation observed after the initial FRC merging can be explained by a larger effective viscosity present in the kinetic simulations due to ion finite Larmor radius effects. For $S^*\sim 20$ and assumed low electron temperature, the thermal ion Larmor radius in the external field, $\rho_i$, is comparable to $d_i$, therefore the ratio of the FRC 'minor' radius to the average ion Larmor radius is of the order $(R_s-R_0)/\langle\rho_i \rangle \approx 3$.

\section{Conclusions}
\label{sec:conclusion}

Both MHD and hybrid simulations show that the magnetic compression via an increasing mirror magnetic field is beneficial for efficient merging of the FRCs.
Without the compression, the FRC merging is very sensitive to initial parameters including the separatrix beta, separatrix radius, $x_s=R_s/R_c$,  and the separation of the FRCs.
Larger values of these parameters are found to be generally unfavorable for the merging.
For example, increasing the separatrix beta from $\beta_s=0.2$ to 0.3 for $x_s\sim0.7$ resulted in failure to merge, and the FRCs eventually drifting apart.
The sensitivity to these parameters is explained by the finite plasma density present in the open field line region between the FRCs.
In addition, is has been shown that without the compression, the FRCs with larger elongation ($E \agt 3$) and a relatively large separatrix radius ($x_s\sim0.7$) merge partially forming a doublet in both the MHD and kinetic simulations.
Formation of the doublet FRC could be related to an increase in the elongation after the FRC merging, and the possible tearing instability in the configurations with large $E$ and $x_s$. 
The FRCs with smaller initial separatrix radius and elongation can merge completely, forming a single-null FRC in the MHD model, and nearly completely merged configuration in the kinetic simulations.

In all merging simulations without the compression the resulting configuration has the separatrix radius and value of the kinetic parameter, $S^*$ comparable to that of the initial FRCs.
The FRC elongation after the merging was increased by a factor $\sim 2.3$ in the cases of partial merging (doublet FRC).
For the complete merging, the elongation increase is found to be by a factor of $\sim 1.7$.
Considering that the tilt mode stability favors smaller values of $S^*/E$ parameter, the FRC merging is beneficial for the tilt stability.

Hybrid simulations performed for $S^*=20-25$ show that the global dynamics of the FRC merging is generally similar to that of the MHD, but the local structure of reconnecting magnetic field corresponds to the Hall reconnection.
In both larger and smaller $x_s$ cases considered, the FRC merging was less complete in hybrid simulations compared to the corresponding MHD cases.

The compression by the increasing mirror magnetic fields at the ends of the simulation region results in faster and more complete merging of the FRCs. 
In that case, the FRCs were fully merged on a time scale of $t\sim20-25t_A$ in both the MHD and hybrid simulations (Figs.\ref{fig:mhd3} and~\ref{fig:hybrid3}).
This is in contrast with the merging results without the compressions performed for the same initial FRC parameters (Figs.\ref{fig:mhd1} and~\ref{fig:hybrid1}).
The complete merging is related not only to the increasing force pushing the FRCs towards the midplane, but also due the change in the external field profile, which has a strong effect on the tearing stability.
Due to a relatively fast compression, the FRCs are shown to get over-compressed and oscillate axially before merging completely.
For values of FRC kinetic parameter considered here, $S^*  \sim 25$, the 2D global dynamics of the FRCs during compressional merging is found to be very similar in the kinetic and MHD simulations.

%\textcolor{red}{Do we need to add something about how it relates to Helion?}
The results of the 2D merging simulation studies presented herein demonstrate that FRCs can merge fully in the presence of compressing mirror fields, and the midplane magnetic reconnection can occur quickly and completely.
%It is expected that 3D simulations are to be performed and compared to the 2D results in the very near future. 
The merging process converts FRC kinetic energy and magnetic field energy near the midplane into ion thermal energy. The simulations shed some light into the operation of Trenta, and build confidence in Helion's technique applied to future device designs. 
The additional physics insight will lead to improved operations of Polaris and  will feed into full device modeling in the near future.
Three dimensional kinetic simulations addressing, in particular, FRC stability during merging and compression are being conducted, and the results will be reported in a separate publication.

\pagebreak
\section*{Acknowledgments}

This research used resources of the National Energy Research Scientific Computing Center (NERSC), a Department of Energy Office of Science User Facility using NERSC award FES-ERCAP0029042. 
This work was supported by the U.S. Department of Energy under contract number DE-AC02-09CH11466 and Department of Energy INFUSE program, grant number 2722. 
The United States Government retains a non-exclusive, paid-up, irrevocable, world-wide license to publish or reproduce the published form of this manuscript, or allow others to do so, for United States Government purposes.

\section*{Data Availability Statement}

The data that support the findings of this study are available from the corresponding author upon reasonable request.

\pagebreak
\bibliographystyle{style/ans_js}                                                                           %custom ANS journal submission template bibliography style
\bibliography{bibliography}

\end{document}